\documentstyle[12pt,aasms4]{article} 
\def\etal{{et~al.}\ }

\def\vol#1  {{{#1}{\rm,}\ }}

\def\etal{et al.\ }

\def\clock{\count0=\time \divide\count0 by 60
     \count1=\count0 \multiply\count1 by -60 \advance\count1 by \time
     \number\count0:\ifnum\count1<10{0\number\count1}\else\number\count1\fi}

\begin{document}
\title{Accuracy of Mesh Based Cosmological Hydrocodes: Tests and Corrections}
\author{Renyue Cen and Jeremiah P. Ostriker}
\vskip 0.5cm
\centerline{Princeton University Observatory, Princeton, NJ 08544}
\vskip 0.5cm
\centerline{(cen,jpo)@astro.princeton.edu}

\begin{abstract}
We perform a variety of tests to
determine the numerical
resolution of the cosmological TVD eulerian code developed
by Ryu \etal (1993).
Tests include $512^3$ and $256^3$ simulations
of a $P_k\propto k^{-1}$ spectrum to check 
for self-similarity and 
comparison of results with those
from higher resolution SPH and grid-based calculations
(Frenk \etal 1998).
We conclude that in regions where density
gradients are not produced by shocks
the code degrades resolution with a Gaussian
smoothing (radius) length 
of $1.7$ cells.
At shock caused gradients (for which
the code was designed)
the smoothing length is $1.1$ cells.
Finally, for $\beta$ model fit clusters,
we can approximately correct numerical resolution by the transformation
$R^2_{core}\rightarrow R^2_{core}-(C\Delta l)^2$,
where $\Delta l$ is the cell size and $C=1.1-1.7$.
When we use these corrections on our previously 
published computations for the SCDM and $\Lambda$CDM models
we find luminosity weighted, zero redshift,
X-ray cluster core radii of $(210\pm 86, 280\pm 67)h^{-1}$kpc,
respectively,
which are marginally consistent  
with observed (Jones \& Forman 1992) values of 
$50-200h^{-1}$kpc.
Using the corrected core radii,
the COBE normalized SCDM model predicts the number
of bright $L_x>10^{43}$erg/s clusters too high
by a factor of $\sim 20$ and
the $\Lambda$CDM model is consistent with observations.
\end{abstract}

\keywords{Cosmology: large-scale structure of Universe 
-- hydrodynamics
-- numerical method}

\section{Introduction}
Two types of hydrodynamic codes are currently
in use for cosmological applications:
mesh based codes pioneered by Vishniac and co-workers (Chiang, Ryu, \&
Vishniac 1989; Ryu, Vishniac, \& Chiang 1990) and
Cen and co-workers (Cen \etal 1990; Cen 1992)
and used by several groups including a TVD 
(``Total Vaviation Diminishing") variant (Ryu \etal 1993)
or a PPM (``Piecewise Parabolic Method") variant (Bryan \etal 1995),
and, alternatively,
particle based smooth particle hydrodynamics codes (``SPH")
used by a variety of groups (Evrard 1988; Hernquist \& Katz 1989; 
Steinmetz 1996; Owen \etal 1998a).
The latter codes can concentrate computational resources
into the highest density regions of greatest interest,
but they suffer in low density regions, at caustics,
and, due to the large computational overhead, they have relatively
small particle number and hence have relatively poor
mass resolution which can induce two body relaxation even in the
high density regions (\cite{kms96}; \cite{sw97}).

But the mesh codes also have, along with their virtues of accurate
treatment of shocks and caustics, good mass resolution, known accuracy
and convergence properties etc, quite serious weaknesses,
the primary among them being poor
spatial resolution in the high density regions.
A detailed comparison of five codes -- three independent mesh codes
and two independent SPH codes 
comparing the virtues and details of the two approaches
was presented in Kang \etal (1994b).
Another major such comparative project
was completed recently (Frenk \etal 1998) 
with a still wider range of codes being tested.

What is the accuracy of a mesh code in resolving structures comparable
to or smaller than the mesh size?
A quantitative assessment of this in the cosmological context was presented
by Cen (1992) for an aerodynamics-based cosmological hydrodynamic code,
which has an effective artificial viscosity of known properties.
Anninos \& Norman (1996) did some very interesting
convergence tests on X-ray clusters by varying 
numerical resolution using a multi-grid eulerian hydrocode.
Bryan \& Norman (1998) examined 
resolution effect on various quantities 
related to simulated X-ray clusters using
PPM eulerian code (Bryan \etal 1995).
Owen \etal (1998b) have studied various scaling 
properties in scale-free ($P_k\propto k^{-1}$), adiabatic SPH simulations.
In the present paper we examine the TVD shock capturing code
originally developed by Harten (1984)
and reformulated with gravity for high mach number,
 cosmological applications
by Ryu \etal (1993).
This code has been used by Cen, Ostriker and co-workers to study
the properties of X-ray clusters of galaxies and Lyman alpha clouds,
and is being used for work on galaxy formation.
The primary result, which we find,
can be stated simply.
In general, the code smooths structure with a gaussian filter 
$e^{-r^2/2 \sigma_r^2}$
such that $\sigma_r = \alpha \Delta l$ where
$\Delta l$ is the cell size and $\alpha$ is the number which
we are interested in fixing through empirical experiments.
Smoothing separately in the three
directions,
$\sigma_r^2=\sigma_x^2+\sigma_y^2+\sigma_z^2$.
Alternatively phrased, an object of true gaussian radius 
$r_{true}$ will have computed radius $r_{comp}$
\begin{equation}
r_{comp}^2 = r_{true}^2 + r_{res}^2
\end{equation}
\noindent where
\begin{equation}
r_{res} = 1.18 \sigma_r; \quad\quad \sigma_r\equiv\alpha\Delta l
\end{equation}
\noindent where the coefficient $1.18$ comes from the
fact that our fitted radius (i.e., core) is defined at
a location where the density drops to half
the central value.
An important, new finding from the current study is that
the TVD shock-capturing code has different 
resolutions in different regions.
We find that the code has a resolution of 
$\sim 1.1$ cells (i.e., $\alpha=0.95$) near shock fronts,
while its resolution in non-shocking, high density regions is lower than
in the shock fronts, $\alpha=1.4$.
Since the scheme has been optimized for 
capturing shocks (rather than, for example,
contact discontinuities),
we should not be surprised by this variation.
The paper is organized as follows:
\S 2 describes the computations to derive
the empirical resolution of the code,
\S 3 presents an application of the results
to previous published simulations using the TVD code
and \S 4 gives conclusions.

\medskip 
\medskip 
\section{Computations}
\medskip 

An extremely difficult aspect of this problem is to design a test,
with a known,
analytically computable solution, that is also
sufficiently realistic to have a bearing on the problems of astrophysical
interest:
in this case the properties of X-ray clusters.
N. Kaiser and also S. White have pointed out to us
that, if the initially assumed spectrum of density perturbations were 
a power law,
with $P_k=Ak^{-1}$ being the most appropriate choice,
then for $\Omega_0=1$, some strict
scaling relations must hold in a perfect simulation.
Specifically, if one
were to look at a given population, e.g.,
the most massive $10\%$ of the bound objects in the universe,
then (see Kaiser 1986) in an adiabatic calculation,
their characteristic sizes should
scale as $(1+z)^{-2}$ and their average temperatures 
scale as $(1+z)^{-1}$.
In our recent simulations of various specific models for the
growth of structure we did not find that these scaling laws
were very well satisfied.
Examining Figures (11) and (12) of Kang \etal (1994a; KCOR hereafter)
we see that the expected scaling law for the temperature is satisfied
to sufficient accuracy (given the
observed statistical fluctuations due to the relatively
small computed sample of clusters),
but the cluster radius evolution is significantly less steep than
is expected.
There are a variety of potential explanations for these
facts.
Three of the most plausible ones are as follows.

\noindent 1. The actual spectrum of the studied CDM model is
not fit by $P_k\propto k^{-1}$ with sufficient precision to make 
self-similarity an expected outcome.

\noindent 2. Resolution corrections
due to numerical inaccuracy are redshift dependent and account for the
departure from the expected scaling.

\noindent 3. The displayed sample was chosen to be of fixed luminosity:
$L_x(0.5<E<4.5keV) > 10^{43}$erg/s,
which is not a sample defined in a scale free way.

To see which, if any, of these explanations is true,
and to better enable us to make appropriate resolution corrections,
we computed two new simulations of a power law
spectrum $P_k=Ak^{-1}$ of
initial density perturbations
with ($512^3$ cells, $256^3$ particles)
and ($256^3$ cells, $128^3$ particles), respectively.
A simulation box of size of $80h^{-1}$Mpc comoving is
used in both simulations,
giving a cell size and nominal resolution of 
($156~h^{-1}$kpc comoving, $312~h^{-1}$kpc comoving),
respectively.
Other parameters, in the familiar notation, were chosen to be 
$\sigma_8=1$, $h=0.5$, $\Omega_{CDM}=0.95$, $\Omega_b=0.05$ to
correspond well to prior work. 
To ensure that the ``truth" remains the same 
in the two simulations, the initial realizations in the two simulations
are exactly the same, with the power spectrum 
being cut off at the Nyquist frequency of the $256^3$ box
for both simulations
(A smooth but rather sharp filter,
$\cos[\pi k/2 k_{nyq,256}]^{1/4}$,
is applied to the power law
spectrum to minimize real space oscillations but maintain
the power law slope as closely as we can).

Figure 1 shows the redshift dependence of the temperature 
(equally weighted) of the absolutely brightest
clusters defined in scale-free fashion: 
the most massive clusters (within a radius of $1.0h^{-1}$Mpc comoving)
which contain 20\% of total mass in the universe at each epoch.
The temperature of each cluster is the X-ray emission-weighted average
over the indicated sphere.
Note that the selection method of clusters at different redshifts 
used here is somewhat different 
from that used in KCOR, which was not scale-free: only the bright clusters
with luminosity $L_x(0.5<E<4.5keV) > 10^{43}$~erg/s 
were selected at each redshift.
Nevertheless
one sees a behavior of the temperature of the set of brightest clusters 
qualitatively similar to what was shown in KCOR (Figure 13 in KCOR):
in both cases the temperature scales with redshift 
approximately as expected: 
$T_x={\rm const.} (1+z)^{-1}$.

We plot, in Figure 2,
$r_{100}$ versus redshift.
Here, $r_{100}$ is the average radius of 
top 20\% (in mass) clusters in each model at each redshift
within which the average density of each cluster 
is $100\bar\rho(z)$ 
[$\bar\rho(z)$ is the global mean at $z$].
As $r_{100}$ is much larger than 
the cell size at all times shown,
resolution effects should be minimal.
We see that the agreement 
between the simulation and the analytic prediction (Kaiser 1986) 
is satisfactory.

Now let us turn to the core radii.
These are much smaller than $r_{100}$
and may be unresolved in our simulation.
Figure 3 shows the redshift dependence of the average cluster core radius 
for the same set of clusters (top 20\% in mass).
Each cluster core is found by fitting the simulated cluster 
emissivity profile to the following equation
\begin{equation}
j = {j_0\over [1+(r/r_{core})^2]^2}
\end{equation}
\noindent As in KCOR (see Figure 12 there) 
we 
see that the cluster core radius
does not scale with redshift as predicted analytically. 
Comparison shows that 
the departure from the expected scaling is as great
for the power law model as for the real CDM-like spectrum, indicating
that spectral curvature is not an important factor over
the redshift range ($0<z<1$) considered.
Furthermore, the scaling behaviors are similar 
when we select the clusters in this powerlaw simulation 
using the same criterion as in KCOR.
Thus, both explanations (1) and (3) are false
and it is likely that the problem is due to 
the redshift dependence of numerical resolution.

Let us now examine the ansatz mentioned in the introduction.
We fit the computed core radius $r_{comp}$ (in comoving units)
with an equation of the form
\begin{equation}
r_{comp}^2 (z)= r_{true}^2(z)+1.39\alpha^2(z)(\Delta l)^2
\end{equation}
The first term on the right hand side represents
the true core size for the {\it actual} model computed
and the second term is $r_{res}^2$,
where $\Delta l$ is the comoving cell size of a simulation.
There are two variables to be solved at
each epoch.
Since the two simulations have identical
initial conditions, $r_{true}$ should be the same
in the two simulations for clusters selected in the same,
scale-free way.
This allows us to solve the above equation for $\alpha(z)$ at each
redshift and then $r_{true}(z)$, both of which
are displayed in Figure 4.
Two points are immediately evident.
First, the resolution of the simulation is
indeed {\it redshift dependent}, as seen in the dependence of $\alpha$
on $z$, ranging from $0.95\pm 0.05$ at redshift one to a
$1.40\pm 0.05$ at redshift zero.
Second, the derived ``true" core size, $r_{true}$, at first sight,
strongly disagrees with the naive analytical expectation,
which states $r_{true}\propto (1+z)^{-\gamma}$ (where
$\gamma$ is a constant thought to be $\sim 1.0$).
Both of these points deserve a thorough understanding.

We address the second point first.
First we note that the assumed power law spectrum has no 
characteristic scale and presumably would give $r_{true}\rightarrow 0$.
However, the actual simulation 
does not possess a perfect $k^{-1}$ power law spectrum.
In fact, the actual input power spectrum to the simulation
is $k^{-1}\cos (\pi k/2 k_{nyq,256})^{1/4}$ for 
$k\in [0.0785,10.053]~h~$Mpc$^{-1}$ comoving 
(where the lower limit is due to the limited box size
and the upper limit is, $k_{nyq,256}$,
the Nyquist frequency for the $256^3$ simulation box)
and zero otherwise.
Figure 5 shows 
the linear r.m.s. density 
fluctuations as a function of top-hat comoving radius 
for the actual smoothed, truncated power law spectrum (solid curve),
which is used in the simulations.
An ideal, untruncated $k^{-1}$ power law spectrum would have the 
the fluctuation spectrum as indicated by the dashed line.
We see that the truncated power law spectrum 
introduces a natural turnover scale around $0.2-0.3h^{-1}~$Mpc
in the density fluctuation spectrum.
Therefore, in the actual simulations under consideration, 
a core can only develop {\it at a size $\sim 0.2-0.3h^{-1}~$}Mpc
{\it or greater.} 
This explains why the derived true core size shown in Figure 4
is constant (within the small noise) from $z=1$ to $z=0$,
simply because 
the true core size for
an untruncated power law spectrum at $z\sim 0$ (with the 
adopted normalization of $\sigma_8=1$)
either happens to be $\sim 0.25h^{-1}~$Mpc or 
is still
smaller than $\sim 0.20-0.30h^{-1}~$Mpc.
Thus, at all higher redshifts,
the derived ``true" core size represents
what is introduced due to the 
truncation of the power, resulting in a nearly constant
core size over the redshift range examined here.

As a consequence, 
a more conservative approach is possible
to obtain a bound on our resolution (i.e., on $\alpha$).
If we assume that the true core 
radius is zero at all redshift, i.e., $r_{true}=0$,
then the measured core radius is entirely due to 
finite numerical resolution
(either in the initial conditions or
in the subsequent hydrodynamic simulations).
We find a redshift dependent bound on the resolution:
at $z=1$, $\alpha< 1.2-1.7$
and $z=0$, $\alpha< 1.5-2.0$.

Let us now turn to the first point:
is the derived value of $\alpha$ and its 
redshift dependence reasonable?
It is not hard to explain results with regard to this.
A shock-capturing code such as the one examined here
is designed to resolve shocks.
In fact, the code is shown to be able to
resolve a shock in about 1-2 cell (top-hat) (Ryu \etal 1993),
which is consistent with what is found here in 
resolving early clusters since at these early times
the regions which dominate X-ray emission
are just undergoing shocking.
On the other hand, the code is also known to
be able to resolve contact discontinuities or non-shocking,
large density gradients
at a lower resolution of 2-3 cells (top-hat) (Ryu \etal 1993).
This is again in agreement with the found resolution
for clusters at lower redshifts ($\sim 1.0-2.0~$cells [Gaussian]),
where shocks are far outside of cluster centers 
and cumulative diffusion with time tends to smooth
the high density central cluster regions.

As a final and quite significant check we show, in Figure 6, the result
taken from Frenk \etal (1998),
where the solid dots are the density profile (spherically averaged)
of a cluster in a controlled volume of a CDM universe
computed by the same TVD code used here with
$N=512$ cells and nominal resolution (i.e., cell size)
of $62.5h^{-1}$kpc.
The open circles and 
solid curve represent a fit to the average profile 
of all the simulations from Frenk \etal (1998),
which is dominated by a few highest resolution simulations 
in the inner region ($r<0.1h^{-1}$Mpc).
The dashed curve is the smoothed profile
of the solid curve by a gaussian with
$\sigma_r=1.65\Delta l$ (i.e., with $\alpha=1.65$).
We see the result computed in the core regions
by our code does in fact
correspond well to a gaussian smoothed
version of the true density profile if the
smoothing length is taken to be $1.65$ cells.
This particular simulated cluster is probably more advanced 
than any cluster in the current simulations.
Therefore, a larger value of $\alpha$ is entirely to be expected,
but is consistent with the bound on $\alpha$ obtained above.

\medskip 
\medskip 
\section{Applications}
\medskip 

Let us now apply the derived results on core radii
to our previous computations of X-ray clusters of galaxies
which had a box size of $85h^{-1}$Mpc and a cell size of $315h^{-1}$kpc.
From Figure 6b of Kang \etal (1994) for the $\Omega_0=1$
SCDM model we have obtained the luminosity-weighted average 
core radius at each redshift, $r_{core,comp}$.
Then, we use equation (1) to compute $r_{core,true}$ at
each redshift, given $\alpha$ as shown in Figure 4. 
Since this SCDM model has comparable amplitudes of the density
fluctuations on the relevant scales 
and comparable abundance of clusters of galaxies
compared to the power law model tested here,
it seems appropriate
to directly use $\alpha$ as shown in Figure 4.

In order to make meaningful assessments we need to 
have an estimate of error on the derived $\alpha$.
We obtain the error on $\alpha$ by finding 
individual $\alpha$ for each pair of clusters found
in the two different resolution simulations.
For clusters selected in a self-similar way as indicated above,
we find that 4 out of 4 clusters in the high resolution
simulation have the counterparts in the low resolution simulation
at $z=0$,
5 out of 5 at $z=0.3$,
6 out of 6 at $z=0.5$,
6 out of 9 at $z=1.0$,
and 12 out of 15 at $z=2.0$.
We do not include clusters that are not paired in the two simulations
in computing the errors on $\alpha$.
We find the $1\sigma$ statistical error 
of $\alpha$ to be $(0.15,0.097,0.049,0.048,0.061)$
at redshift $z=(0.0,0.3,0.5,1.0,2.0)$, respectively,
with the dispersion being $(0.25,0.19,0.11,0.11,0.20)$.
The identification of each pair of clusters 
is unambiguous with 3-d r.m.s displacement being
less than one simulation cell at all epochs examined.

If a cluster's average velocity dispersion with
some large radius (a few times the core radius)
is fixed and the emissivity profile is assumed to be that as in
equation (3),
then one can show that approximately
$L_x\propto r_{core}^{-1}$.
Further assuming that the luminosity function
has a slope of $-2$, i.e., $n(>L)\propto L^{-2}$,  
as indicated by both simulations and observations (see
Figures 1-4 of  Kang \etal 1994),
we are able to correct the number of bright X-ray clusters.
Figure 7 shows the computed core radii, the corrected core radii,
the computed number of X-ray clusters brighter
than $L>10^{43}$ erg/sec,
and corrected number of X-ray clusters brighter
than $L>10^{43}$ erg/sec,
in the SCDM model, from redshift zero to one.
The errorbars on $r_{core,corr}$ and $n_{corr}$
are obtained by propagating errorr through
the following equations: 
$\Delta r_{core,corr}/r_{core,corr}=\Delta \alpha/\alpha$,
$L_x\propto r_{core}$
and
$n(>L_x) \propto L_x^{-2}$
(see below for a discussion on errors).
The most significant result from this exercise
is that the apparent positive evolution of bright
X-ray clusters previously found in the SCDM model
seems due to the fact that the lower redshift clusters
are relatively more underresolved.
Correcting this redshift-dependent resolution effect
seems to show that the bright clusters
are consistent with no evolution (or weak evolution) up to redshift one,
in better agreement with observations and
semi-analytic studies (Henry \etal 1992).
Figure 8 shows the same results for the $\Lambda$CDM
model (Cen \& Ostriker 1994).
Since the $\Lambda$CDM model
is significantly different from the power law model computed
here, it is somewhat tricky as to how to apply $\alpha$ derived
here to the $\Lambda$CDM model.
We make the following observation.
Since a $\sigma_8\sim 0.5$ power law model
has approximately the same 
cluster abundance as the $\Lambda$CDM model (e.g., Cen 1998),
it seems most appropriate
to apply the $\alpha$ at $z=1$ in the power law model
to clusters at $z=0$ in the $\Lambda$CDM model.
For clusters in the $\Lambda$CDM model at higher redshift
we simply use our best estimates of $\alpha$ by extrapolation.
Note that the corrected zero redshift luminosity
weighted X-ray core radii in the (SCDM,$\Lambda$CDM)
models are $(210\pm 45, 280\pm 60)h^{-1}$kpc,
respectively.
This errorbars on $r_{true}$ 
are estimated based on the errors on $r_{comp}$.
This is to be compared with observations
by Jones \& Forman (1992) of $50-200h^{-1}$kpc.

For both models we see that our previous computations
have overestimated the core radii by factors of $1.7-3.1$
and 
underestimated the number of bright clusters
by a varying factor from about 3 to 10.
We were aware of the resolution issue when we wrote
Kang \etal (1994) and Cen \& Ostriker (1994)
and thus treated the computed numbers
of bright clusters
as lower bounds to the true numbers.
Thus, the present exercise
has the primary effect of strengthening
our previous conclusion that the COBE
normalized CDM model overpredicts the number of 
bright X-ray clusters by a very large factor ($\sim 20$).
The $\Lambda$CDM model, 
revised to include corrections described
here, would be approximately
consistent with observations
[note that we found a plotting error in our previous
published results: the vertical values of the simulated results
in Figure 1 of both Kang \etal (1994) and Cen \& Ostriker (1994)
are too large by a factor of $\ln (10)=2.3$;
our revised statement above with regard to the number of bright
clusters in the two models includes the
correction of this error].

Finally, 
let us estimate the systematic errors associated with the corrected
luminosity of a cluster using the resolution correction method 
described here.
Assuming that $L_x\propto r_{core}^{-1}$
and taking the form $L_{true}=(D\pm \Delta D)L_{comp}$,
we have $D=r_{comp}/r_{true}=\sqrt{1+1.18\alpha \Delta l/r_{true}}$.
Taking the $z=0$ solid square in Figure 7 as an example
(which has largest extrapolation among our results)
with $\alpha=1.4$ and $r_{true}=0.65\Delta l$, 
we have $D=1.9$;
i.e., the correction (due to systematic error)
on the X-ray luminosity of clusters
at $z=0$ is as large as the computed value,
it thus appears that 
systematic errors associated with extrapolated X-ray 
cluster luminosity is still very large for the published
simulations.
But, we note that, 
if we trust the derived values of
$r_{true}$,
then in new simulations at a dynamic range 
of $768^3$ now achievable with the same box size
($L=85h^{-1}$Mpc) as the previously published ones,
the resolution correction would be small
with $D=1.38$, i.e., 38\% correction (systematic error),
and relatively reliable.
The associated errorbar on $D$ (statistical error)
would be $\Delta D/D=\Delta r_{comp}/r_{comp}
=\Delta \alpha/\alpha=0.15/1.4=10.7\%$ for 
the clusters in the SCDM model at $z=0$.
The clusters in the $\Lambda$CDM model at $z=0$,
$(D,\Delta D)$ would be $(1.20,0.05)$.
Hence a $95\%$ percent upper bound (including both
systematic and statistical errors) would have 
an upward correction on computed value of only $1.97$ 
and $1.50$, respectively, for clusters at $z=0$ in the new
simulations of SCDM and $\Lambda$CDM models.
Corrections for clusters at higher redshifts
would be still smaller.
A larger simulation box would diminish
the statistical errorbars,
while higher resolution would further reduce
systematic errorbars.

\medskip 
\medskip 
\section{Conclusions}
\medskip 

To summarize our results,
we find an effective gaussian
smoothing length of approximately 1.7 cells except
in regions where the density
gradients are caused by shocks
(for which the TVD code is optimized)
where the smoothing length is approximately $1.1$ cells.
Density profiles can
be deconvolved with the smoothing
length when the correction is small:
$R^2_{core,true} = R^2_{core,comp}-(C\Delta l)^2$,
where $C=1.1-1.7$.
But results are not to be trusted
if the computed core radii of
clusters are less than $1.1\Delta l$.

Applying the derived resolution
effect to our previous X-ray cluster simulations
we find that our previous computations
underestimate the number of bright clusters
by a varying factor from about 3 to 10.
We estimate that the error on the corrected clusters luminosities 
are still very large, thus the correction is not reliable.
In addition, the redshift evolutions of bright clusters
in the models are altered to varying extent.
Our previous conclusion that the COBE
normalized CDM model overpredicts the number of 
bright X-ray clusters by a very large factor 
is greatly strengthened.
Finally, we note that, 
with new simulations at a dynamic range 
of $768^3$ now achievable with the same box size
($L=85h^{-1}$Mpc) as the previously published ones,
the resolution correction would be small
[$(38\%,20\%)$, respectively]
and relatively reliable.

We are happy to acknowledge 
support from grants NAGW-2448, NAG5-2759, AST91-08103
and ASC93-18185 and useful conversations 
with L. Hernquist, N. Kaiser, U. Pen and S. White.
It is a pleasure to acknowledge Pittsburg Supercomputing Center
for allowing us to use Cray C90 supercomputer.
We would like to thank the hospitality of ITP during our stay
when this work is completed, and the financial support from ITP
through the NSF grant PHY94-07194.
Some of the computation was performed
at the Princeton SGI Origin 2000
which is supported by grant  from the NCSA Alliance Center.

\vfill\eject

\newpage

\figcaption[Figure 1]{
shows the redshift dependence of the temperature 
(equally weighted) of the brightest clusters defined in scale-free fashion: 
the most massive clusters (within a radius of $1.0h^{-1}$Mpc comoving)
which contain 20\% of total mass in the universe at each epoch.
The open circles are the results from the $512^3$ cell simulation
and the solid dots are from the $256^3$ cell simulation
with $2\sigma$ 
{\it statistical}
errorbars.
The solid lines are the best Kaiser fits for 
the power law model in question: $T(z) = T_0 (1+z)^{-1}$.
It  is seen that 
the results agree with analytical predictions very well.
\label{fig1}}

\figcaption[Figure 2]{
shows $r_{100}$ as a function of redshift.
Here, $r_{100}$ is the average radius of 
top 20\% (in mass) clusters at each redshift
with which the average density of each cluster 
is $100\bar\rho(z)$ [$\bar\rho(z)$ is the global mean at $z$].
\label{fig2}}

\figcaption[Figure 3]{
shows the redshift dependence of the core radius
(equally weighted) of the same brightest clusters as shown in
Figure 1 (defined in scale-free fashion). 
The open circles are the results from the $512^3$ cell simulation
and the solid dots are from the $256^3$ cell simulation
with $2\sigma$ 
{\it statistical}
errorbars.
The solid curves are the best Kaiser fits for 
the power law model in question in the form,
$r_{core}(z) = r_0 (1+z)^{-2}$.
\label{fig3}}

\figcaption[Figure 4]{
shows
derived $\alpha(z)$ (solid squares)
and $r_{true}(z)$ (open circles) 
as a function of
redshift in the range $z=0-1$.
Here $r_{true}$ is the core corrected radius.
\label{fig4}}

\figcaption[Figure 5]{
shows 
the linear r.m.s. density 
fluctuations as a function of top-hat comoving radius 
for the actual smoothed, truncated power law spectrum (solid curve),
which is used in the simulations.
Also shown as the dashed curve is that for 
an ideal, untruncated $k^{-1}$ power law spectrum without
the small scale power truncation.
\label{fig5}}

\figcaption[Figure 6]{
The open circles and 
solid curve represent a fit to the profile averaged over
all simulations (but dominated by the few highest resolution simulations) 
presented in Frenk \etal (1998).
The dashed curve is the smoothed profile
of the solid curve by a gaussian with
$\sigma_r=\alpha \Delta l$ with $\alpha=1.65$.
We see the result computed in the core regions
by our code does in fact
correspond well to a gaussian smoothed
version of the true density profile if the
smoothing length is taken to be $1.65$ cells.
\label{fig6}}

\figcaption[Figure 7]{
shows the computed core radii, the corrected core radii,
the computed number of X-ray clusters brighter
than $L>10^{43}$ erg/sec,
and corrected number of X-ray clusters brighter
than $L>10^{43}$ erg/sec,
in the SCDM model, from redshift zero to one.
The errorbars are $1\sigma$ {\it statistical}.
\label{fig7}}

\figcaption[Figure 8]{
shows the computed core radii, the corrected core radii,
the computed number of X-ray clusters brighter
than $L>10^{43}$ erg/sec,
and corrected number of X-ray clusters brighter
than $L>10^{43}$ erg/sec,
in the $\Lambda$CDM model, from redshift zero to one.
The errorbars are $1\sigma$ {\it statistical}.
\label{fig8}}

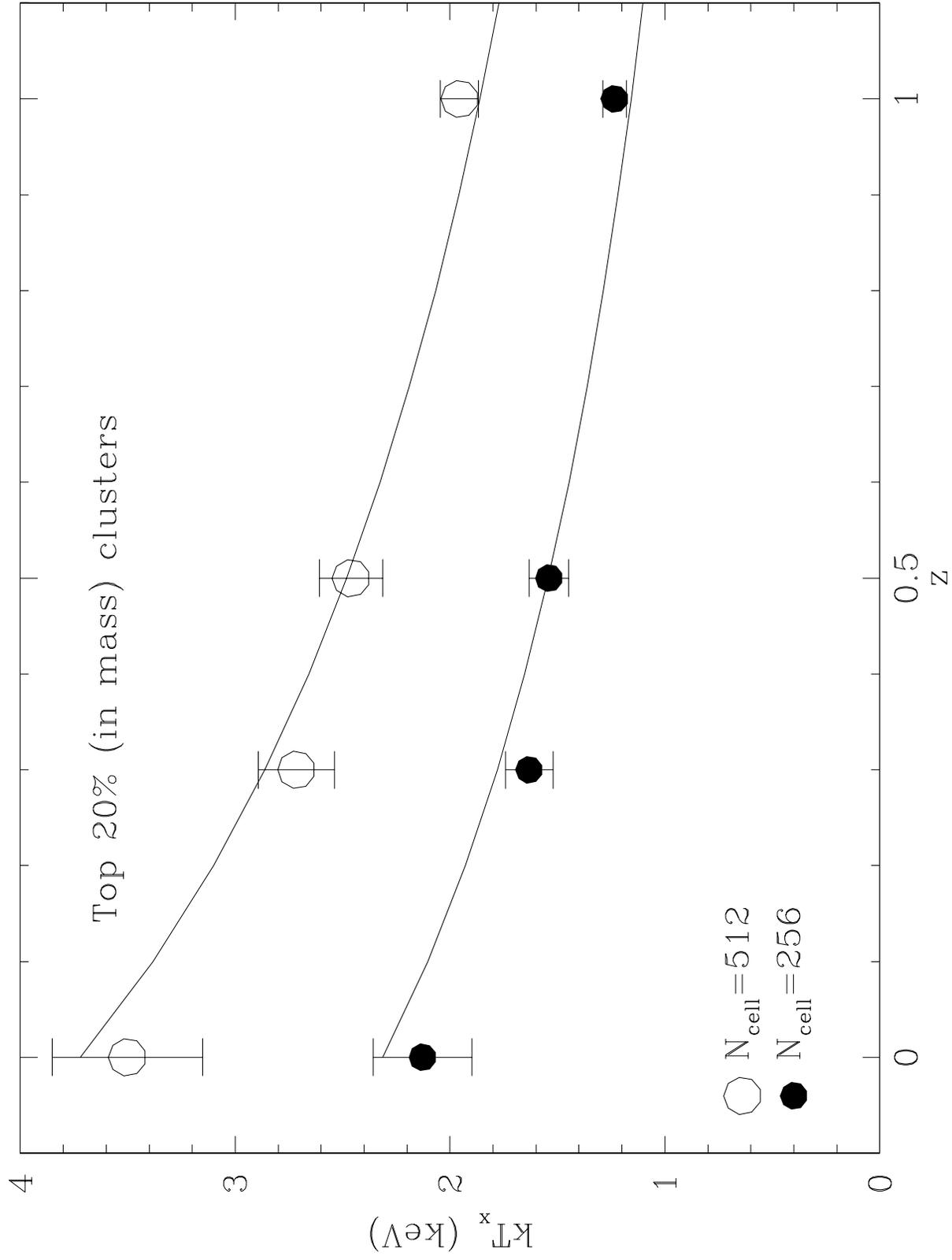
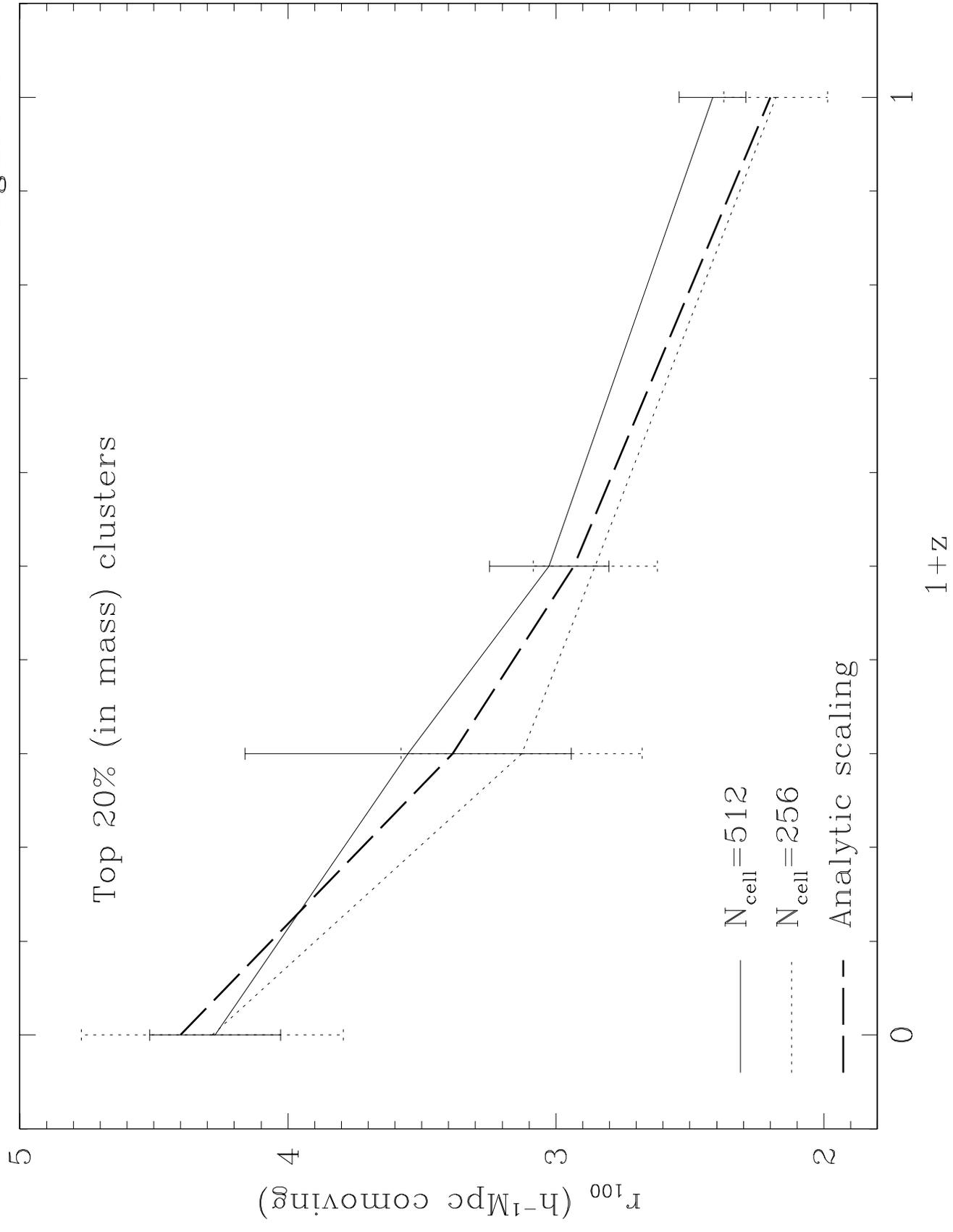
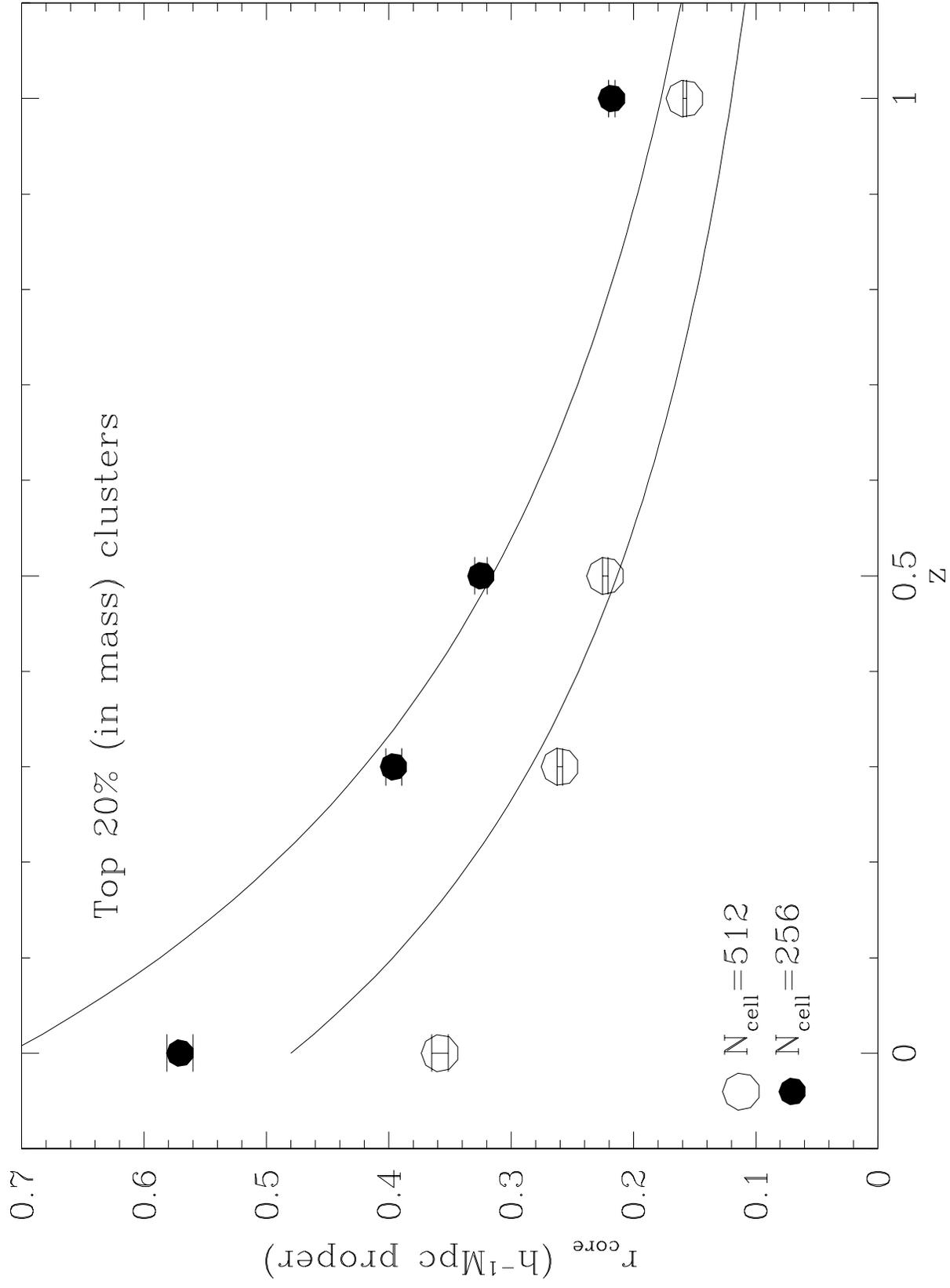
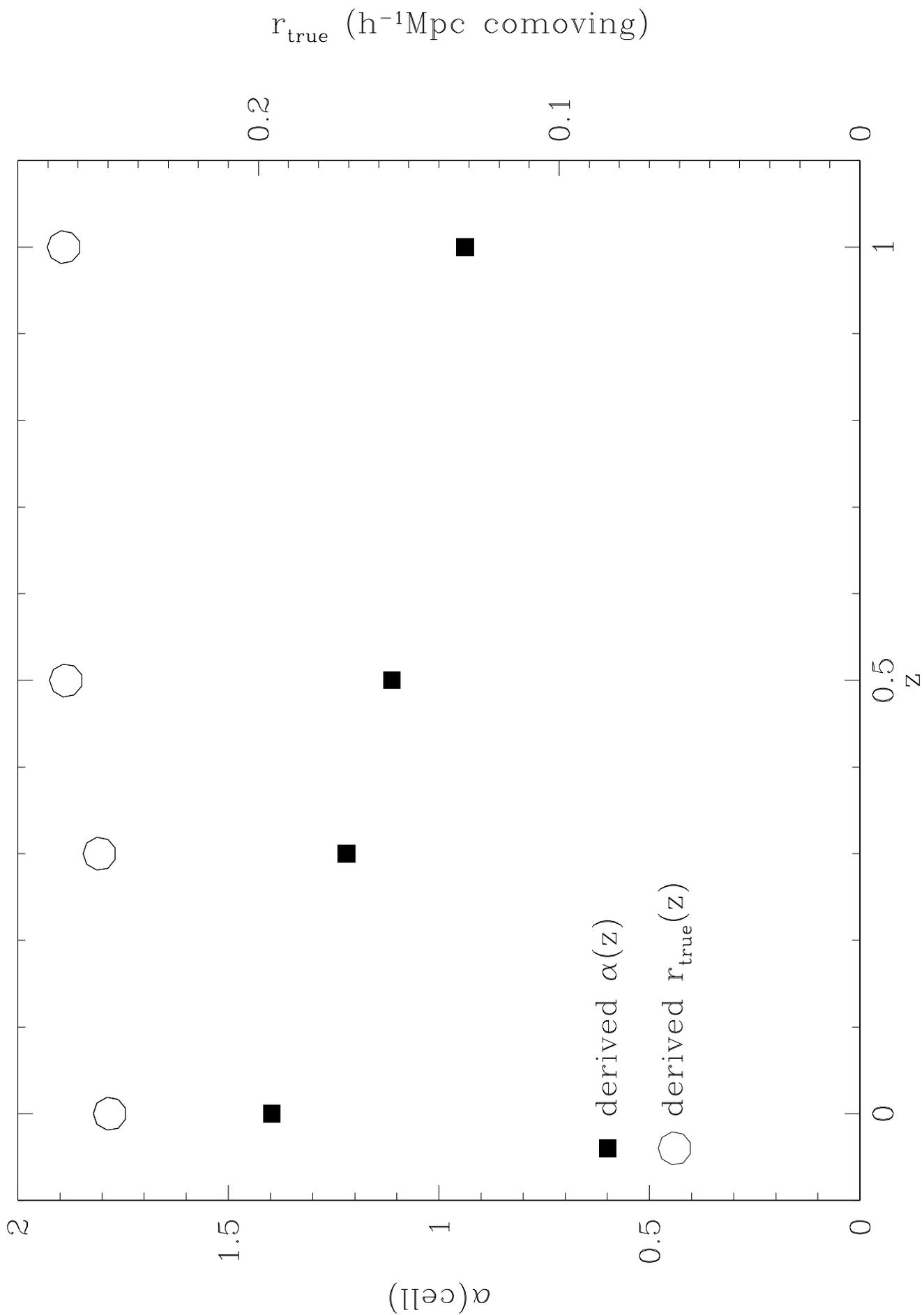
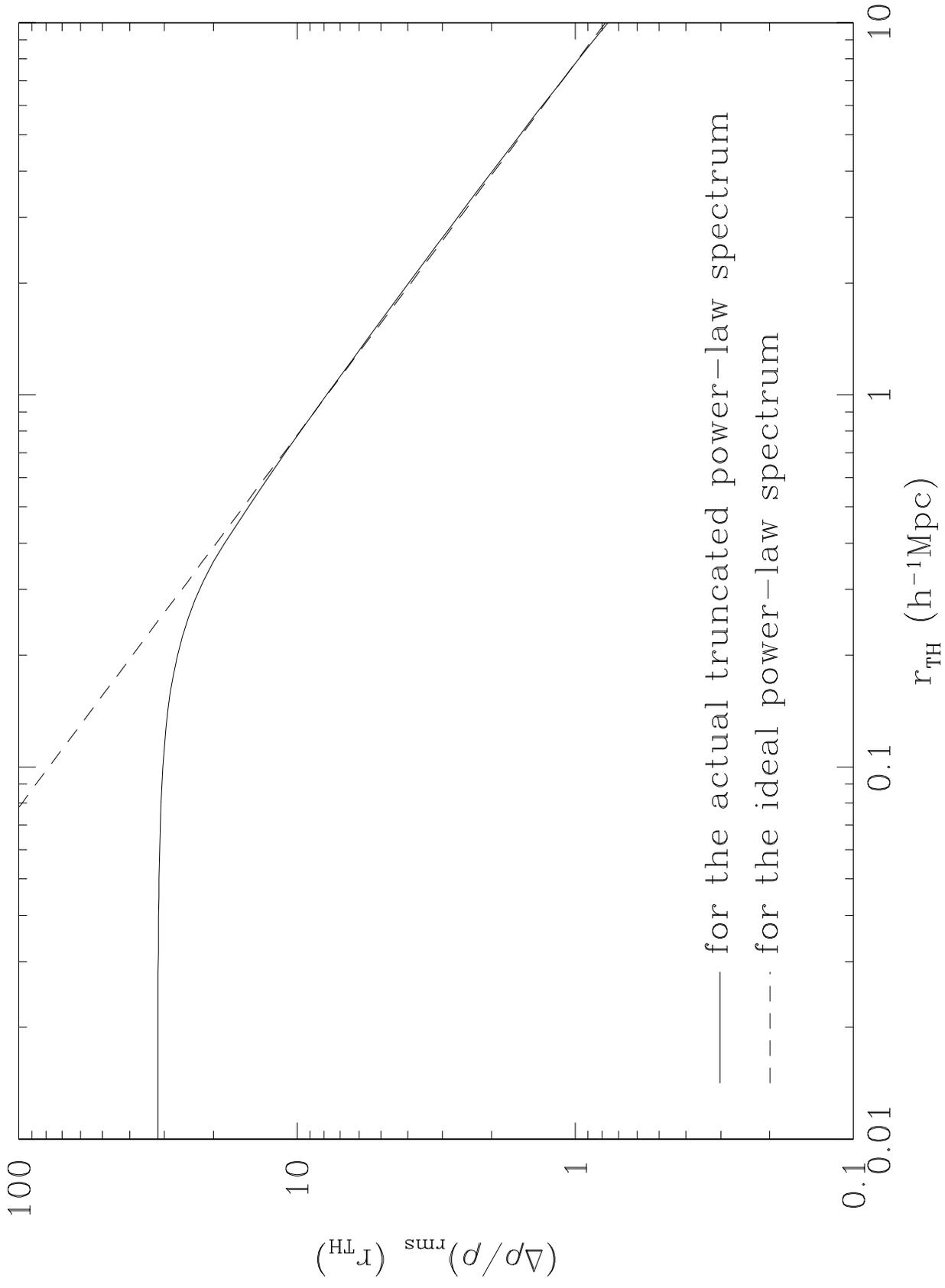
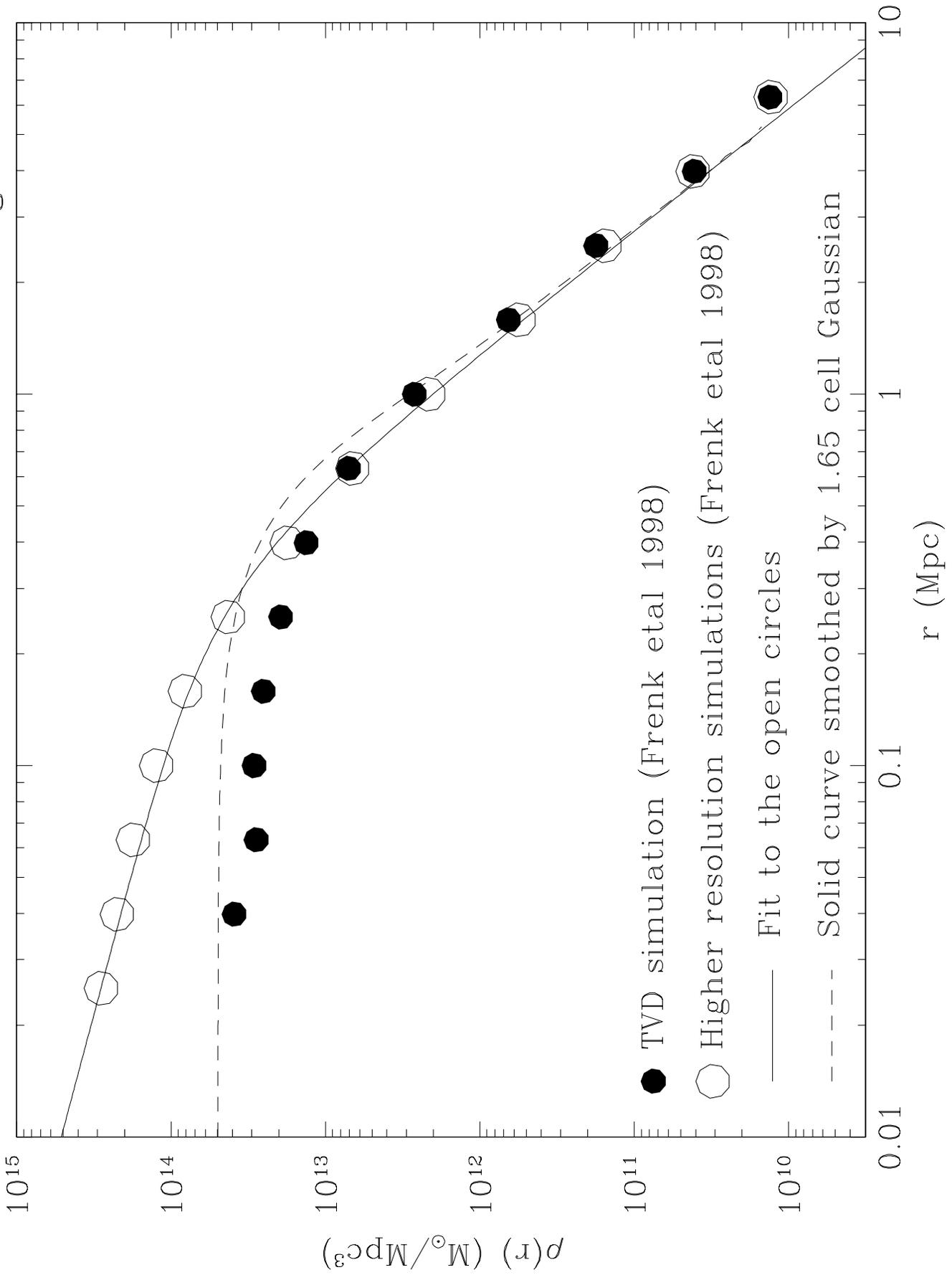
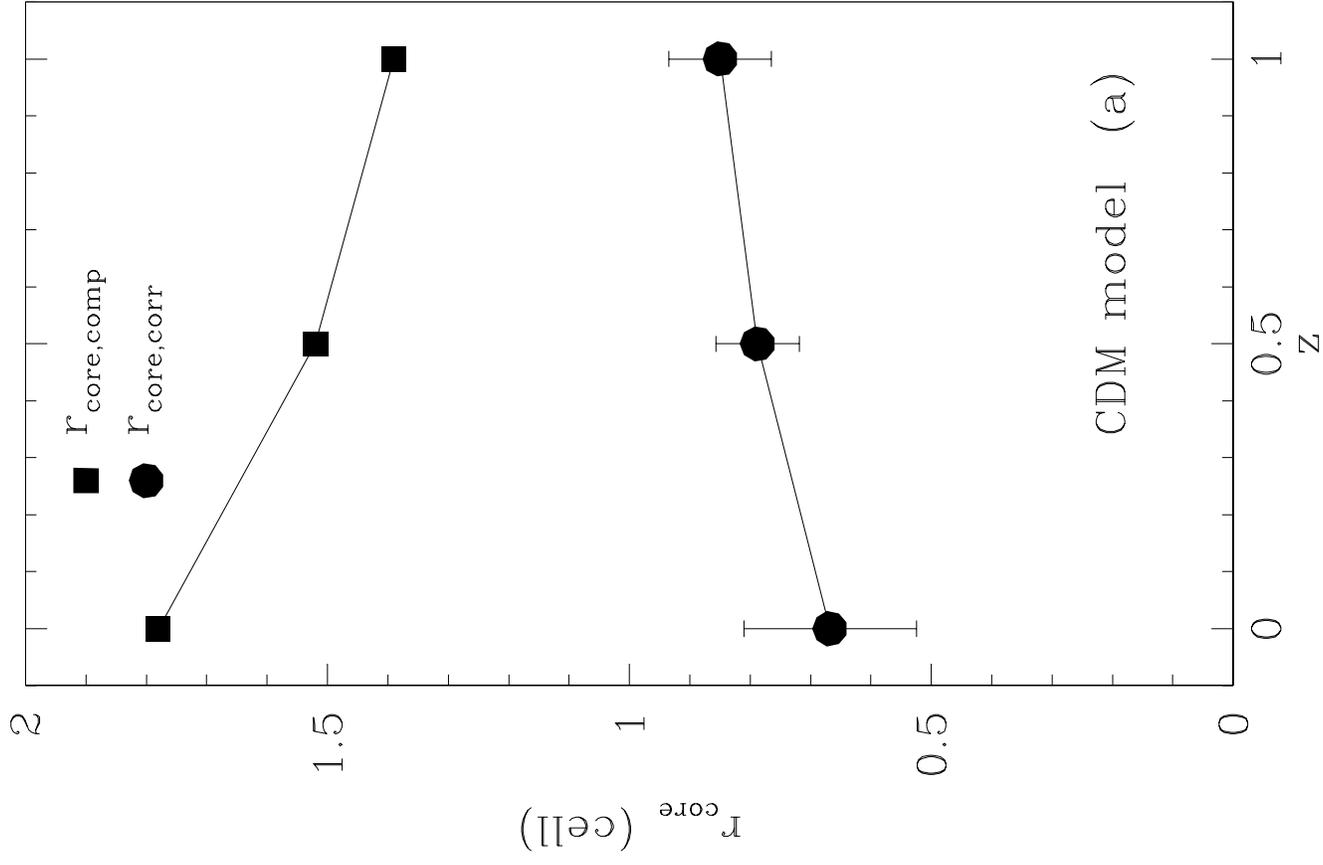
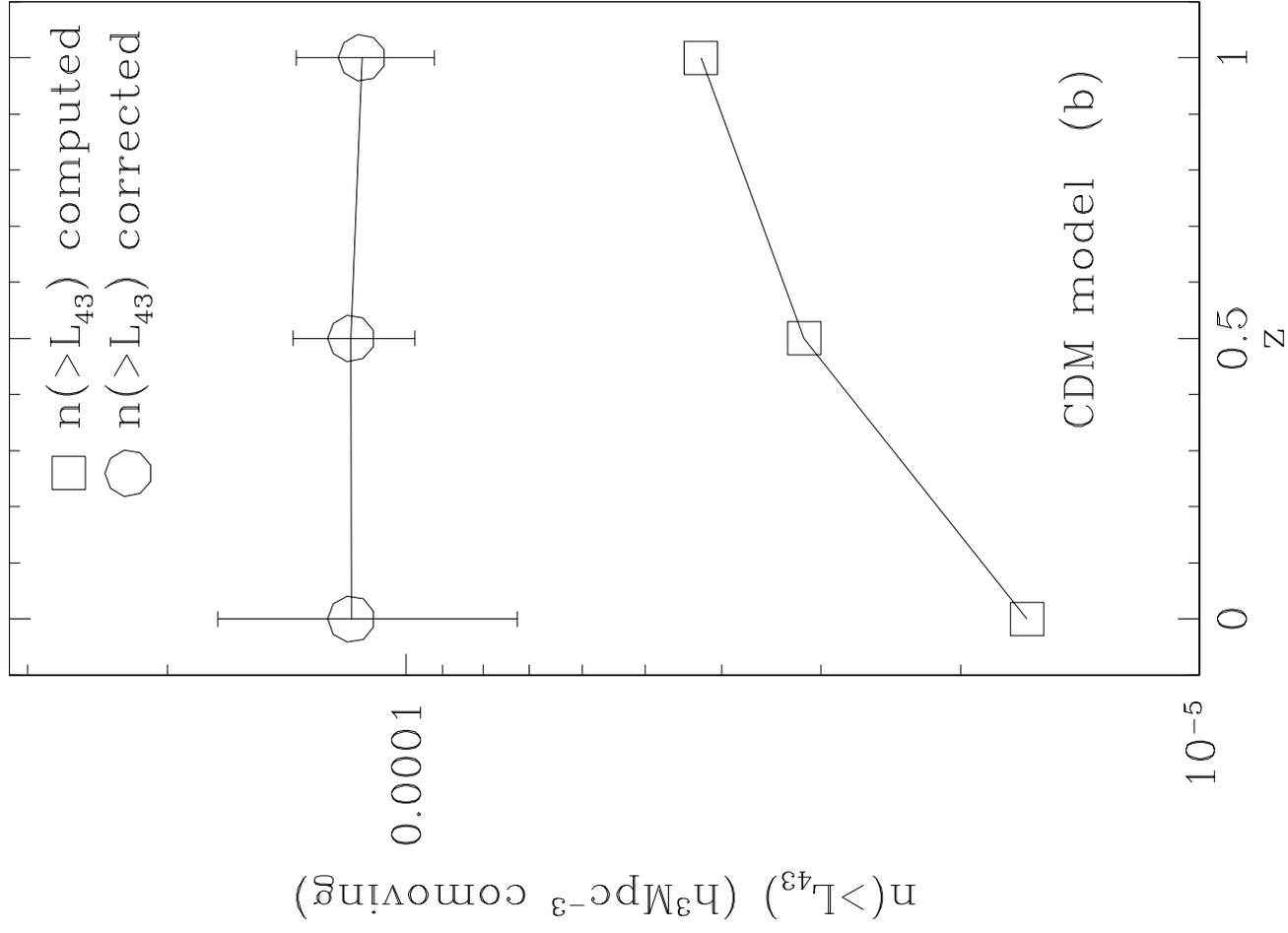
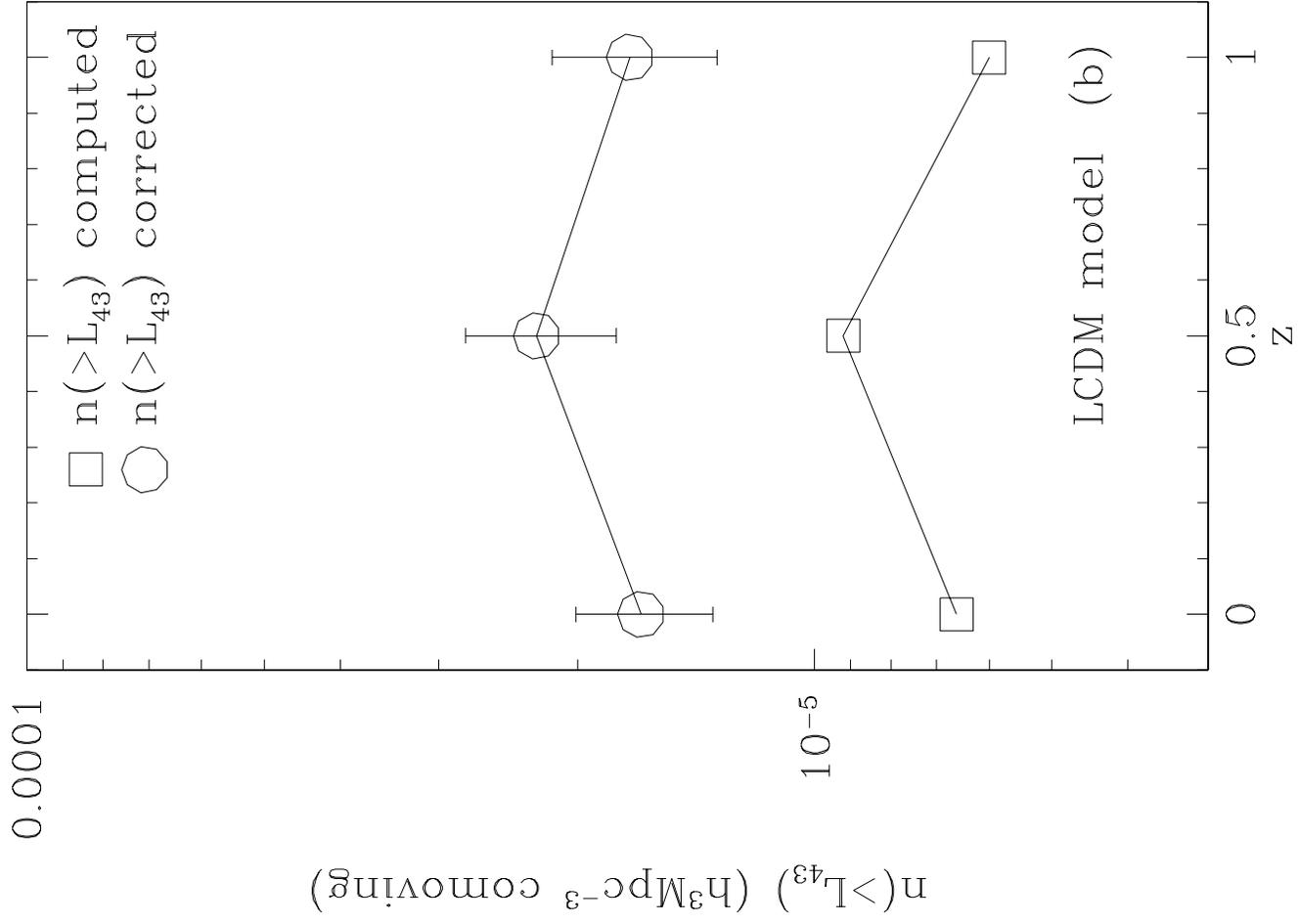
\end{document}